\documentclass[a4paper,12pt]{article}
\usepackage{amssymb}
\usepackage{amsmath}
\date{}
\title{On the String Pair Creation in D$p\,$-D$p'$ Brane System}
\author{A. Jahan\\Department of Physics, Amirkabir University of Technology\\P. O. Box: 15875-4413, Tehran, Iran\\{ jahan@aut.ac.ir}}
\begin{document}
\maketitle
\begin{abstract}
We address the bosonic string pair creation in a system of parallel D$p\,$-D$p'$ ($p<p'$) branes by applying the path integral formalism. We drive the string pair creation rate by calculating the one loop vacuum amplitude of the setup in presence of the background electric field defined over the D$p'$-brane. It is pointed out that just the components of the electric field defined over the $p$ spatial directions (the common directions along which the both D-branes are extended) give rise to the pair creation.
\end{abstract}
\section*{\large 1\quad Introduction}
It is well known that the superstring theory vacuum becomes instable in presence of a constant electric background field and decays into the string pairs quite similar to the so called Schwinger effect in QED  [1-11]. In more recent studies on the issue the possible role of the lower dimensional D-branes are considered on the string pair creation in presence of the constant electric and additional magnetic backgrounds. It has been pointed out that the string pair creation takes place in the D$p\,$-D$p$ and D$p\,$-$\overline{\textrm{D}}p$ systems in type-IIA or type-IIB models, although the criterion for occurrence of the vacuum decay plays more drastic role in the D$p\,$-$\overline{\textrm{D}}p$ system due to some exponential factors depending on the background configurations [5]. Some possible effect of a fixed magnetic background on the vacuum decay has been also considered: a fixed magnetic field can greatly enhance the pair creation rate in the case of a weak electric background and the system undergoes infinite number of phase transitions [10, 11].\\
In the present work we consider a D$p$-brane and a D$p'$-brane on which the bosonic strings end. We assume a constant electric field over the D$p'$-brane. The goal is to calculate the string pair creation rate using the path integral technique for such a setup. It is demonstrated that the stability of vacuum depends on the direction of the electric field and the dependence of the pair creation rate on the background field components is discussed. In fact this system is
extensively studied in [11, 12] for the case of supersrings (PBS saturated setups) where the result of this study is implicitly presented. However there are some differences
between our analyze and those appeared in [11, 12]: First, we engage the apparatus of path
integral to unveil the underlying physics of the system. Second, we consider a relatively simple setup by
confining our analyze to the bosonic strings. Third, we employ the one-loop open string formalism while the tree level closed string approach (boundary state formalism) is exploited in [11]. \\ We review the path integral derivation of the bosonic degrees of freedom satisfying different boundary conditions in the next section. In section 3, the vacuum amplitude of the setup in presence of a constant electric field is derived by utilizing the path integral formalism and the the possible implications of background field direction on the pair creation rate is discussed. Throughout this work we assume the Euclidean signature for both of the world-sheet and target space manifolds. However continuation to the Lorentzian signature is assumed for the space-time manifold after integration over the bosonic degrees of freedom.
\section*{\large 2\quad One Loop Vacuum Energy: Zero Background}
Let us begin with the bosonic string action in the $d$ dimensional space-time interacting with a constant gauge field background through its end points
{\setlength\arraycolsep{2pt}
\begin{eqnarray}\label{1}
S_B&=&\frac{T}{2}\int d^2\sigma\partial_aX^\mu\partial_a X^\mu+S_{gh}[b,c]+S_{int}.
\end{eqnarray}}
We postpone the explicit form of the interaction action $S_{int}$ until the section 3. At one loop level the annulus and m\"{o}bius strip diagrams are the only diagrams relevant to the open string partition function coupled to a U(1) background. Since the m\"{o}bius strip has one boundary its contribution to the one loop vacuum amplitude is associated with a configuration in which the string bears the same charges at its end points. We shall assume a setup in which the string interacts just with one end. So the only diagram pertinent to our calculations is the annulus diagram. We begin with the path integral form of the free energy [6-8]
{\setlength\arraycolsep{2pt}
\begin{eqnarray}\label{2}
\mathcal F=\int_0^\infty \frac{ds}{s}\int^\prime DX^\mu DbDc\,e^{-S_0[X]-S_{gh}[b,c]-S_{int}}.
\end{eqnarray}}
provided that $X^\mu(\sigma,\tau)=X^\mu(\sigma,\tau+s)$. The prime over the second integral means that the contribution of zero modes are excluded in evaluating the above path integral. In presence of a constant electric background the free energy gets an imaginary part and the vacuum begins to decay into the superstring pairs with decay rate given by
\begin{equation}\label{3}
w=-2\textrm{Im}\mathcal F.
\end{equation}
For a string ending on two different D-brans with a relative distance $Y$ there are $1+p$ degrees of freedom satisfying the Neumann-Neumann (NN) boundary condition $p'-p$ degrees of freedom satisfying the Neumann-Dirichlet (ND) boundary condition and $d-p'$ satisfying the Dirichlet-Dirichlet (DD). So, the partition function becomes $ Z_B=Z_{N}^{1+p}Z^{p'-p}_{ND}Z_{D}^{d-p'-1}Z_{gh}$ where the partition function of ghost fields is given by $Z_{gh}=\frac{T}{2s}Z^{-2}_{N}$. The path integral evaluation of the partition function entails the following Fourier expansions for the typical fluctuations $X_{N}$, $X_{D}$ and $X_{ND}$
{\setlength\arraycolsep{2pt}
\begin{eqnarray}\label{4-6}
X_{N}&=&\sum_{m\in\mathbb Z}\sum_{n\in\mathbb N}\chi_{mn}u_{mn},\\
X_{D}&=&\sum_{m\in\mathbb Z}\sum_{n\in\mathbb N^+}\chi_{mn}v_{mn}+\frac{l}{\pi}\sigma,\\
X_{ND}&=&\sum_{m\in\mathbb Z}\sum_{n\in\mathbb N+\frac{1}{2}}\chi_{mn}u_{mn}.
\end{eqnarray}}
$\mathbb N=\mathbb N^+\bigcup\{0\}$. Here, the eigen-modes are $u_{mn}=e^{i\omega_m\tau}\cos n\sigma$ and $v_{mn}=e^{i\omega_m\tau}\sin n\sigma$. Furthermore, we define $\omega_m=m\omega $, where $\omega=\frac{2\pi}{s}$. and $\lambda_{mn}=\frac{1}{4}sT(\omega_m^2+n^2)$. Also we introduce the matrix $\textbf{M}_m^D$ as
\begin{equation}\label{7}
[\textbf{M}_m^D]_{nn'}= \left\{\begin{array}{ll}
2\lambda_{m0}, &n,n'=0\\
\lambda_{mn}\delta_{nn'}, &n,n'\neq0 .
\end{array} \right.
\end{equation}
and $\textbf{M}_m^{D}=\textbf{M}_m^{ND}=\lambda_{mn}\delta_{nn'}$. Therefore, we find the corresponding partition functions by integrating over the fluctuations as
{\setlength\arraycolsep{2pt}
\begin{eqnarray}\label{8}
Z_{N}(s)&=&\int^\prime DX_{N}e^{-S_0[X_{N}]}=\prod_{m\in\mathbb Z}\det{'}(\textbf{M}_m^N)^{-\frac{1}{2}}\\\nonumber
&=&\prod_{m=1}^\infty\frac{1}{2\lambda_{m0}}
\prod_{n=1}^\infty\frac{1}{\sqrt{\lambda_{0n}}}\prod_{m=1}^\infty
\prod_{n=1}^\infty\frac{1}{\lambda_{mn}}\\\nonumber
&=&\sqrt{\frac{T}{2s}}q^{-\frac{1}{24}}\prod^\infty_{n=1}\frac{1}{1-q^{n}}.
\end{eqnarray}}
where $q=e^{-s}$. In the same way, we obtain
{\setlength\arraycolsep{2pt}
\begin{eqnarray}\label{9}
Z_{D}(s)&=&\prod_{m\in\mathbb Z}\det(\textbf{M}_m^D)^{-\frac{1}{2}}
=\prod_{n=1}^\infty\frac{1}{\sqrt{\lambda_{0n}}}
\prod_{m=1}^\infty\prod_{n=1}^\infty\frac{1}{\lambda_{mn}}\\\nonumber
&=&q^{\frac{1}{2\pi}T l^2-\frac{1}{24}}\prod^\infty_{n=1}\frac{1}{1-q^{n}}.
\end{eqnarray}}
Also we find the partition function of the degree of freedom satisfying ND boundary condition as [13]
{\setlength\arraycolsep{2pt}
\begin{eqnarray}\label{10}
Z_{ND}(s)&=&\prod_m\det(\textbf{M}_m^{ND})^{-\frac{1}{2}}
=\prod_{n\in\mathbb N+\frac{1}{2}}\frac{1}{\sqrt{\lambda_{0n}}}\prod_{m=1}^\infty\prod_{n\in\mathbb N+\frac{1}{2}}\frac{1}{\lambda_{mn}}\\\nonumber
&=&q^{-\frac{1}{48}}\prod^\infty_{n=1}\frac{1}{1-q^{n-\frac{1}{2}}}.
\end{eqnarray}}
In arriving at the equations (8), (9) and (10) we have gained the infinite products
{\setlength\arraycolsep{2pt}
\begin{eqnarray}\label{11-12}
\frac{\sinh\pi x}{\pi x}&=&\prod_{n=1}^\infty\Big(1+\frac{x^2}{n^2}\Big),\\
\prod_{n=1}^\infty\frac{1}{an^2}&=&\frac{\sqrt a}{2\pi},
\end{eqnarray}}
together with the zeta-regularization of the infinite sums $\sum_{n=1}1=-\frac{1}{12}$, $\sum_{n=1}n=-\frac{1}{12}$ and
{\setlength\arraycolsep{2pt}
\begin{eqnarray}\label{11-12}
\sum_{n\in\mathbb N+\frac{1}{2}}n&=&\frac{1}{24},\\
\sum_{n\in\mathbb N+\frac{1}{2}}1&=&0.
\end{eqnarray}}
\section*{\large 3\quad Constant Electric Background}
For a string interacting via one end, say $\sigma=0$, the interaction term is
{\setlength\arraycolsep{2pt}
\begin{eqnarray}\label{14}
S_{int}&=&-\frac{i}{2}E_I\int_{\sigma=0} d\tau (X^0\partial_\tau X^I-X^I\partial_\tau X^0).
\end{eqnarray}}
The electric field defined over the D$p'$-brane can be written as $\vec E=\vec E_\bot+\vec E_\|$ where we have defined $\vec E_\bot=(0,\ldots,0,E_{p+1},\ldots,E_{p'})$ and $\vec E_\|=(E_1,\ldots,E_p,0,\ldots,0)$. We combine the free and interaction bosonic actions to introduce the actions
{\setlength\arraycolsep{2pt}
\begin{eqnarray}\label{15-16}
A_{I}&=&S_0[X^0]+\sum_{i=1}^{p'}S_0[X^i]+S_{int},\\
A_{II}&=&\sum_{i'=p'+1}^{d-1}S_0[X^{i'}].
\end{eqnarray}}
Upon introducing $\mathbf{X}=(X^0,\ldots,X^{p'})$ the action $A_I$ can be recast in
\begin{equation}\label{17}
A_{I}=\int d^{\,2}\sigma \mathbf{X}^{\,\scriptsize\textrm{t}}\mathcal M \mathbf{X},
\end{equation}
where the matrix $\mathcal M$ is
\begin{equation}\label{18}
\mathcal M=\frac{T}{2}
\left(\begin{array}{cccc}
\Box&-i\frac{E_1}{T}\delta(\sigma)\partial_\tau&\cdots&-i\frac{E_{p'}}{T}\delta(\sigma)\partial_\tau\\
i\frac{E_1}{T}\delta(\sigma)\partial_\tau&\Box&&\\
\vdots&&\ddots&\\
i\frac{E_{p'}}{T}\delta(\sigma)\partial_\tau&&&\Box
\end{array}\right).
\end{equation}
Integration over the variable $\mathbf{X}$ yields the corresponding partition function as
\begin{equation}\label{19}
Z_I[\vec E]=\prod_{m\in\mathbb Z}\det \mathcal{M}_{m}^{-\frac{1}{2}}.
\end{equation}
Here the matrix $\mathcal M_{m}$ is given by
\begin{equation}\label{20}
\mathcal M_{m}=
\left(\begin{array}{cccc}
\textbf{M}_{m}^0&-\textbf I_m^1&\cdots&-\textbf I_m^{p'}\\
\textbf I_m^1&\textbf{M}_{m}^1&&\\
\vdots&&\ddots&\\
\textbf I_m^{p'}&&&\textbf{M}_{m}^{p'}
\end{array}\right).
\end{equation}
with $[\textbf{I}_m^I]_{nn'}=mE_I$ and $0\leq n,n'<\infty$. Now with the aid of the formulas
{\setlength\arraycolsep{2pt}
\begin{eqnarray}\label{21-22}
\sum_{n=1}\frac{1}{\omega^2_m+n^2}&=&\frac{\pi}{2\omega_m}\coth\pi\omega_m-\frac{1}{2\omega_m^2},\\
\sum_{n=1}\frac{1}{\omega^2_m+(n-\frac{1}{2})^2}&=&\frac{\pi}{2\omega_m}\tanh\pi\omega_m,
\end{eqnarray}}
we obtain
{\setlength\arraycolsep{2pt}
\begin{eqnarray}\label{23}
Z_{I}[\vec E]&=&Z_N^{p+1}Z_{ND}^{p'-p}\prod_{m=1}\frac{1}{1-\frac{\vec E^2_{\perp}}{T^2}-\frac{\vec E^2_{\|}}{T^2}\coth^2\pi\omega_m}\\\nonumber
&=&\frac{T}{4\pi}e^{\frac{3}{s}\pi^2}Z_N^{p-1}Z_{ND}^{p'-p}\sqrt{1-\frac{\vec E^2}{T^2}}\prod_{m=1}^\infty\frac{1}
{(1-e^{-2\pi(\omega_m+\epsilon)})(1-e^{-2\pi(\omega_m-\epsilon)})},
\end{eqnarray}}
where we have used the Jacobi modular property of the Dedekind eta function $\eta(\tau)=e^{i\frac{\pi \tau}{12}}\prod_{m=1}^{\infty}(1-e^{2\pi i\tau m})=\frac{\eta(-\frac{1}{\tau})}{\sqrt{-i\tau}}$ to write
\begin{equation}\label{24}
\prod_{m=1}^\infty\big(1-e^{-2\pi\omega_m}\big)^2=\frac{T}{4\pi}e^{\frac{3}{s}\pi^2}Z_N^{-2}.
\end{equation}
The parameter $\epsilon$ in (24) is given by
\begin{equation}\label{24}
\epsilon =\frac{1}{2\pi}\ln (\alpha+\sqrt{\alpha^2-1}),
\end{equation}
where
{\setlength\arraycolsep{2pt}
\begin{eqnarray}\label{25}
\alpha&=&\frac{1-\frac{\vec E^2_{\perp}}{T^2}+\frac{\vec E^2_{\|}}{T^2}}{1-\frac{\vec E^2_{\perp}}{T^2}-\frac{\vec E^2_{\|}}{T^2}}
=\frac{1+\cos2\theta\frac{{\vec E}^2}{T^2}}{1-\frac{{\vec E}^2}{T^2}}.
\end{eqnarray}}
In this equation we have introduced the parameterizations $\vec E_{\|}=\vec E\cos\theta$ and $\vec E_{\bot}=\vec E\sin\theta$ through defining the angle $\theta$. The modular property of the Jacobi theta function  ($z=e^{2\pi i\nu}$)
{\setlength\arraycolsep{2pt}
\begin{eqnarray}\label{26}
\Theta_1(\nu|\tau)&=&2q^{\frac{1}{8}}\sin\pi\nu\prod_{m=1}^{\infty}(1-e^{2\pi i\tau m})(1-ze^{2\pi i\tau m})(1-z^{-1}e^{2\pi i\tau m})\\\nonumber
&=&-\frac{e^{-i\frac{\nu^2}{\tau}}}{\sqrt{-i\tau}}\Theta_1\Big(\frac{\nu}{\tau}\Big{|}-\frac{1}{\tau}\Big),
\end{eqnarray}}
provide us with an equivalent representation for (24) of the form
{\setlength\arraycolsep{2pt}
\begin{eqnarray}\label{27}
Z_{I}[\vec E]&=&T\frac{E_\|}{4\pi}
\frac{ q^{\frac{1}{2\pi}\epsilon^2-\frac{1}{12}}}{\sin(\frac{s\epsilon}{2})}Z_N^{p-1}Z_{ND}^{p'-p}
\prod_{n=1}^\infty(1-q^{n+i\epsilon})^{-1}(1-q^{n-i\epsilon})^{-1},
\end{eqnarray}}
where we have used the identity $\sinh^2 x=\frac{1}{2}\big(\sqrt{1+\sinh^22x}-1\big)$ to achieve
\begin{equation}\label{28}
\sinh \pi\epsilon=\frac{E}{T}\frac{
\cos\theta}{\sqrt{1-\frac{\vec E^2}{T^2}}},\qquad E=|\vec E|.
\end{equation}
As a criterion for the accuracy of our calculations we look at the fate of the above equation in $E\rightarrow 0$ limit. To this end let's note that in weak field limit we have $\epsilon\simeq\frac{1}{\pi}\frac{E}{T}\cos\theta$. So, as is expected, in this limit one attains $Z_I[0]=Z_{N}^{1+p}Z^{p'-p}_{ND}$. There are divergences associated with equation (29) at the singularity points given by $s_n=\frac{2\pi n}{\epsilon}$. These singularities located along the integration contour give rise to an imaginary part which can be extracted with the help of the well-known formula
\begin{equation}\label{29}
\frac{1}{x-i\varepsilon}=\textrm{P}\frac{1}{x}+i\pi\delta(x).
\end{equation}
where P stands for the Cauchy principal value.
The total bosonic partition function will be $Z_B[\vec E]=Z_{I}[\vec E]Z_{II}Z_{gh}$ with $Z_{II}=Z_{D}^{d+p'-1}$. Therefore from (3) and (29) we obtain the pair creation rate as
\begin{equation}\label{30}
w=\frac{E_\|}{\epsilon}\sum^\infty_{n=1}(-1)^{n+1}e^{-\epsilon n}Z_B\Big(\frac{2\pi n}{\epsilon}\Big).
\end{equation}
Now the appearance of $E_\|$ in (29) and (32) makes its role on the pair creation more apparent. Indeed for $\vec E_\|=0$ and $\vec E_\bot\neq 0$ the decaying of vacuum into the string pairs vanishes. This refers to the fact that in this case we have $\alpha=1$ which implies $\epsilon=0$ in (29). In other words when $\vec E_\|=0$ the vacuum is stable and the partition function becomes $Z_B[\vec E_\bot]=Z_B[0]V_{p'}\sqrt{1-\frac{\vec{E}^2_\bot}{T^2}}$ where $V_{p'}$ is the D$p'$-brane volume. This observation leads us to conclude that the pair creation can not take place in a set up in which the electric field of the form $\vec E=\vec E_\bot$ is assumed over the D$p'$-brane.
\section*{\large Conclusion}
Using the path integral formalism we re-derived the free energy of a system of D$p\,$-D$p'$ brane with a constant electric field considered over the D$p'$ brane volume. We obtained the string pair creation rate for this set up and showed that the string pairs tunnel out from the vacuum provided that the electric field has non-zero components along the directions which  the both D-branes are extended.
\section*{\large Appendix}
For the matrix $\mathcal{O}$ defined as $\mathcal{O}=\mathcal O_{1}+\mathcal O_{2}$ with
\begin{equation}\label{45}
\mathcal{O}_{1}=
\left(\begin{array}{ccc}
\textbf{A}_0&&\\
&\ddots&\\
&&\textbf{A}_k
\end{array}\right),\qquad
\mathcal{O}_{2}=
\left(\begin{array}{cccc}
0&-c_1\textbf I&\cdots&-c_k\textbf I\\
c_1\textbf I&0&&\\
\vdots&&\ddots&\\
c_k\textbf I&&&0
\end{array}\right).
\end{equation}
we can write its determinant as ${}$
\begin{equation}\label{46}
\det\mathcal O=\det\mathcal{O}_{1}e^{\scriptsize{\textrm{Tr}}\ln(1+\mathcal{O}^{-1}_{1}\mathcal{O}_{2})}.
\end{equation}
Therefore from $\ln(1+x)=-\sum_{n=1}\frac{(-1)^n}{k}x^n$ and $\textrm{Tr}(\textbf{A}_i\textbf{I})^{n}=(\textrm{Tr}\textbf{A}_i)^{n}$ one obtains
\begin{equation}\label{47}
\textrm{Tr}(\mathcal{O}^{-1}_{1}\mathcal{O}_{2})^{2n}=2(-1)^n\Big[\textrm{Tr}\textbf{A}_0^{-1}\Big(\sum_{i=1}^kc^2_k
\textrm{Tr}\textbf{A}^{-1}_i\Big)\Big]^n,
\end{equation}
which leaves us with
\begin{equation}\label{47}
\det\mathcal{O}=\det \textbf{A}_0\prod_{i=1}^k\det \textbf{A}_i\Big[1+\textrm{Tr}\textbf{A}_0^{-1}\Big(\sum_{i=1}^kc^2_i
\textrm{Tr}\textbf{A}^{-1}_i\Big)\Big].
\end{equation}
\section*{\large References}
[1]\hspace{0.2cm}C. P. Burgess, Nucl. Phys. \textbf{B294} 427 (1987).\\\
[2]\hspace{0.2cm}C. Bachas, M. Porrati, Phys. Lett. \textbf{B296} 77 (1992).\\\
[3]\hspace{0.2cm}C. Bachas, Phys. Lett. \textbf{B374} 37 (1996).\\\
[4]\hspace{0.2cm}A. A. Bytsenko, S.D. Odintsov and L. Granda, Mod. Phys. Lett. \textbf{A11} 2525 (1996).\\\
[5]\hspace{0.2cm}M. Porrati, arXiv:hep-th/9309114. \\\
[6]\hspace{0.2cm}C. Acatrinei, Nucl. Phys. \textbf{B539} 513 (1999). \\\
[7]\hspace{0.2cm}C. Acatrinei, R. Iengo, Phys. Lett. \textbf{B482} 420 (2000).\\\
[8]\hspace{0.2cm}J. Ambjorn, Y. M. Makeenko, G. W. Semenoff and R. J. Szabo, JHEP \textbf{0302} 026 (2003). \\\
[9]\hspace{0.2cm}J. Schwinger, Phys. Rev \textbf{D82} 664 (1951).\\\
[10]\hspace{0.2cm}J. H. Cho, P. Oh, C. Park and J. Shin,
JHEP \textbf{0505} 004 (2005).\\\
[11]\hspace{0.2cm}J. X. Lu, Shan-Shan Xu, JHEP \textbf{0909} 093 (2009). \\\
[12]\hspace{0.2cm}B. Chen, X. Liu, JHEP \textbf {0808} 034 (2008).\\\
[13]\hspace{0.2cm}J. Fr\"{o}ehlich, O. Grandjean, A. Recknagel and V. Schomerus, Nucl. Phys. \textbf{B583} 381 (2000).
\end{document}